\documentclass[manuscript]{aastex}
\usepackage{fixltx2e}
	\usepackage{float}

\shorttitle{Energy release in the lower solar atmosphere of a solar flare}
\shortauthors{Sharykin and Kosovichev}

\begin{document}

\title{FLARE ENERGY RELEASE IN THE LOWER SOLAR ATMOSPHERE NEAR THE MAGNETIC FIELD POLARITY INVERSION LINE}

\author{I.N. Sharykin\altaffilmark{1,2}, V.M.~Sadykov\altaffilmark{3},
A.G. Kosovichev\altaffilmark{3}, S.~Vargas-Dominguez\altaffilmark{4},
I.V.~Zimovets\altaffilmark{1}}

\affil{Space Research Institute of RAS, Moscow 117997, Russia}

\altaffiltext{1}{Space Research Institute (IKI) of the Russian Academy of Sciences}
\altaffiltext{2}{Institute of Solar-Terrestrial Research (ISTP) of the Russian Academy of Sciences, Siberian Branch}
\altaffiltext{3}{New Jersey Institute of Technology}
\altaffiltext{4}{Universidad Nacional de Colombia}


\begin{abstract}

We study flare processes in the solar atmosphere using observational data for a M1-class flare of June 12, 2014, obtained by New Solar Telescope (NST/BBSO) and Helioseismic Magnetic Imager (HMI/SDO). The main goal is to understand triggers and manifestations of the flare energy release in the photosphere and chromosphere using high-resolution optical observations and magnetic field measurements. We analyze optical images, HMI Dopplergrams and vector magnetograms, and use Non-Linear Force-Free Field (NLFFF) extrapolations for reconstruction of the magnetic topology \textbf{and electric currents}. The NLFFF modelling reveals interaction of two magnetic flux ropes with oppositely directed magnetic field in the PIL. These flux ropes are observed as a compact sheared arcade along the PIL in the high-resolution broad-band continuum images from NST. In the vicinity of PIL, the NST H$\rm\alpha$ observations reveal formation of a thin three-ribbon structure corresponding to a small-scale photospheric magnetic arcade. The observational results evidence in favor of location of the primary energy release site in the chromospheric plasma with strong electric currents concentrated near the polarity inversion line. In this case, magnetic reconnection is triggered by the interacting magnetic flux ropes forming a current sheet elongated along the PIL.

\end{abstract}

\section{INTRODUCTION}

Magnetic reconnection is believed to be the main mechanism of solar flare energy release (see \cite{Priest2000,Priest2002} and references therein). The most popular standard CSHKP flare model \citep{Carmichael1964,Sturrock1966,Hirayama1974,Magara1996,Tsuneta1997} assumes that the reconnection occurs in a quasi-vertical current sheet in the solar corona beneath of an upward moving plasmoid. In the framework of this model the primary energy release (magnetic reconnection in the current sheet) and electron acceleration take place in the low-density corona. It is likely that a large fraction of eruptive solar flares follows this scenario \citep[e.g.][]{Liu2008,Krucker2008,Fletcher2011}. Obviously, non-eruptive flare events cannot be interpreted in the framework of the standard model. It has been argued that the magnetic reconnection process can be triggered not only in the corona but also in the low, partially ionized, layers of the solar atmosphere \citep[e.g.][]{Georgoulis2002, Chae2003}. Recently developed models of chromospheric magnetic reconnection \citep[e.g.][]{Leake2012,Leake2013,Leake2014,Ni2015} have been applied to relatively small scale phenomena, such as chromospheric jets, and Ellerman bombs. However, it is unclear if magnetic reconnection in the chromosphere may play a significant role in solar flares, and what kind of magnetic topology may be involved in the reconnection process. In this respect, investigation of the flare energy release in the lower solar atmosphere is particularly interesting.

It has been known since observations of \cite{Severnyi1958} that the solar flares appear first in the vicinity of the polarity inversion line (PIL) of the line-of-sight magnetic field, and that the flare emission spreads outside the PIL as the flare develops. Also, he found that a considerable gradient of the magnetic field across the PIL is required for the flare appearance. Further observations provided new knowledge about flare processes near the PIL. For example,\cite{Wang1994}, \cite{Kosovichev2001}, and \cite{Sudol2005} showed that the flare onset is often associated with sharp changes of magnetic field near the PIL. The statistical analysis of SOHO/MDI line-of-sight magnetograms, presented in the work of \cite{Welsch2008} showed that the PILs with strong magnetic field gradient are associated with flux emergence. The most recent statistical study made by \cite{Schrijver2016} illustrated that X-class flares are associated with strong-field, high-gradient polarity inversion lines (SHILs) created during emergence of magnetic flux into active regions. A detailed 3D modelling of the magnetic field presented in the work of \cite{Inoue2016} revealed conditions in the vicinity of PIL, favourable for magnetic reconnection and triggering solar flares. High-resolution observations of the formation of H$\rm_{\alpha}$ ribbons showed that a flux rope elongated along a PIL in the low solar atmosphere becomes unstable following an enhancement of its twists \citep{Wang2015}. Thus, it is especially interesting to study flare processes in the vicinity of PIL, as well as connections with the dynamics of magnetic fields and electric currents.

In this paper we present a new investigation of the M1.0 non-eruptive solar flare of 12~June, 2014, started at around 21:00~UT, peaked at 21:12~UT and ended at $\approx$21:30~UT. It occurred in active region NOAA 12087 located approximately 50 degrees South-East from the disk center. This event was previously studied by \cite{Kumar2015,Sadykov2015,Sadykov2016} who provided a general analyse of the structure and dynamics in the chromosphere and corona, as well as study of the chromospheric evaporation process. Our specific goal is to investigate in detail the structure and variations of the magnetic field and electric currents in the PIL. This event was selected for analysis due to the availability of various unique observations of the solar atmosphere made by Interface Region Imaging Spectrograph, IRIS \citep{DePontieu2014}, New Solar Telescope \citep[NST, ][]{Goode2012} in the Big Bear Solar Observatory (BBSO), the Reuven Ramaty High-Energy Solar Spectroscopic Imager \citep[RHESSI, ][]{Lin2002}, Helioseismic Magnetic Imager \citep[HMI, ][]{Scherrer2012} and Atmospheric Imaging Assembly \citep[AIA, ][]{Lemen2012} instruments onboard Solar Dynamics Observatory \citep[SDO, ][]{Pesnell2012}. This flare showed an intriguing activity in the vicinity of the PIL, which was discussed in the previous works \citep{Sadykov2014,Sadykov2015,Kumar2015}. \cite{Sadykov2014} first reported on observations of a small-scale photospheric flux-rope structure that started untwisting imiadeately after the X-ray peak. \cite{Kumar2015} presented a detailed morphological evidence of magnetic reconnection in the vicinity of the PIL between two small opposite polarity sunspots. Using the NST H$_{\alpha}$ images they concluded that an interaction between two J-shape loops led to formation of a twisted flux-rope along the PIL. They also detected plasma inflows near the PIL, which were recognised as a signature of magnetic reconnection of a flux-rope with the overlaying magnetic field. The IRIS and RHESSI observations of the selected flare were previously discussed in the paper of \cite{Sadykov2015}, who presented a detailed analysis of Doppler shift maps reconstructed for different spectral lines, and compared these maps with the RHESSI X-ray images. It was concluded that the chromospheric evaporation was triggered not only by accelerated particles but also by heat flux from the hot coronal plasma. \cite{Sadykov2016} studied the relationship between the chromospheric evaporation flows deduced from IRIS observations and the magnetic field topology of the flare region. Using vector magnetograms from HMI they reconstructed the magnetic field topology and the quasi-separatrix layer (QSL), and found that the upflow (blueshift) regions as well as the H-alpha flare ribbons are magnetically connected to the PIL region via a system of low-lying loop arcades (with height $\lesssim$4.5 Mm). This allowed them to propose an interpretation of the chromospheric evaporation based on the geometry of local magnetic fields, and the primary energy source associated with the PIL. However, the processes in the PIL itself were not previously studied. This is a topic of our paper.

In addition to analysis of the H$\rm_{\alpha}$-line wing filtergrams and broadband TiO images (7057~\AA) with high spatial resolution, which correspondingly show processes in the lower chromosphere and the photosphere, we use nonlinear force-free magnetic field (NLFFF) modelling to explain some topological peculiarities and energy release of the flare. In particular, the NLFFF modelling can help to find possible sites of magnetic reconnection, and to understand where the current sheet is formed. Magnetic reconnection and flare triggering can be associated with shearing motions, magnetic flux emergence or cancellation. To detect the plasma flows and compare them with other observational data we use the HMI Dopplergrams.

Thus, the focus of this paper is a detailed analysis of physical processes in the vicinity of the PIL in the low atmosphere during the solar flare using analysis of the electric current system, evolution of magnetic field topology and dynamics of plasma flows. The paper is divided into four sections. Section 2 describes the observational data. Section 3 is devoted to analysis of NST images, electric currents and magnetic field topology (using the HMI data). Discussion and conclusions are presented in Section 4.


\section{OBSERVATIONAL DATA}

To investigate the fine spatial structure of the flare energy release site we use $H_{\alpha}$ data from Visible Imaging Spectrometer (VIS) as well as TiO filter (7057~\AA) images from Broadband Filter Imager (BFI), obtained at the 1.6 m New Solar Telescope \citep{Goode2012,Varsik2014} and available from the BBSO web site (http://bbso.njit.edu/). The new system of adaptive optics, the large aperture, speckle-interferometry postprocessing \citep{Woeger2008} \textbf{and good seeing conditions} made it possible to obtain images of the Sun with high spatial resolution up to the diffraction limit. The pixel size of the VIS images is about 0.029$^{\prime\prime}$, which is approximately 3 times smaller than the telescope diffraction limit $\lambda_{H\alpha}/D\approx 0.084^{\prime\prime}$. The pixel size of the BFI TiO images is about 0.034$^{\prime\prime}$ with the diffraction limit $\approx 0.09^{\prime\prime}$. The time cadence between two subsequent $\rm H_{\alpha}$ line scans in 5 wavelength bands (6563 $\rm\AA$ $\pm$1.0, $\pm$0.8, $\pm$0.4 and the line core) is $\approx$15 seconds.

In this paper we do not discuss X-ray and ultraviolet observations in details as it was made earlier by \cite{Sadykov2015} and \cite{Kumar2015}; we focus on the processes near the PIL in the lower solar atmosphere.

For the understanding of the flare energy release it is extremely important to determine the magnetic field topology of the PIL as magnetic reconnection can be triggered in a magnetic field configuration different from the CSHKP model. We study the magnetic field structure using observations of Helioseismic Magnetic Imager \citep{Scherrer2012}. The HMI makes spectropolarometric observations of the magneto-sensitive Fe~I line (6173\AA) by two different cameras in six wavelength filtergrams covering the whole line profile. Calculations of the optical continuum intensity, line-of-sight magnetic field and Doppler velocity with the time cadence of 45 s and the spatial resolution of $\sim 1^{\prime\prime}$ are based on the right and left circular polarization measurements. To obtain the full magnetic field vector with the time cadence of 720 s the HMI also makes measurements of the linear polarization profile. The spatial resolution of vector magnetograms is the same as in the case of the LOS magnetograms. In our data sets, the time cadence of the $H_{\alpha}$ filtergrams from NST is the shortest. Thus, we consider physical processes on time scales longer than 15 seconds.

The 3D structure of magnetic field in the flare region was reconstructed using the NLFFF optimization method \citep{Wheatland2000} available from the package SolarSoft. For the boundary condition we used the disambiguated HMI vector magnetograms \citep{Centeno2014}.

\section{RESULTS}

\subsection{FINE STRUCTURE OF THE PIL REGION}

\begin{figure}[h!]
\centering
\includegraphics[width=0.8\linewidth]{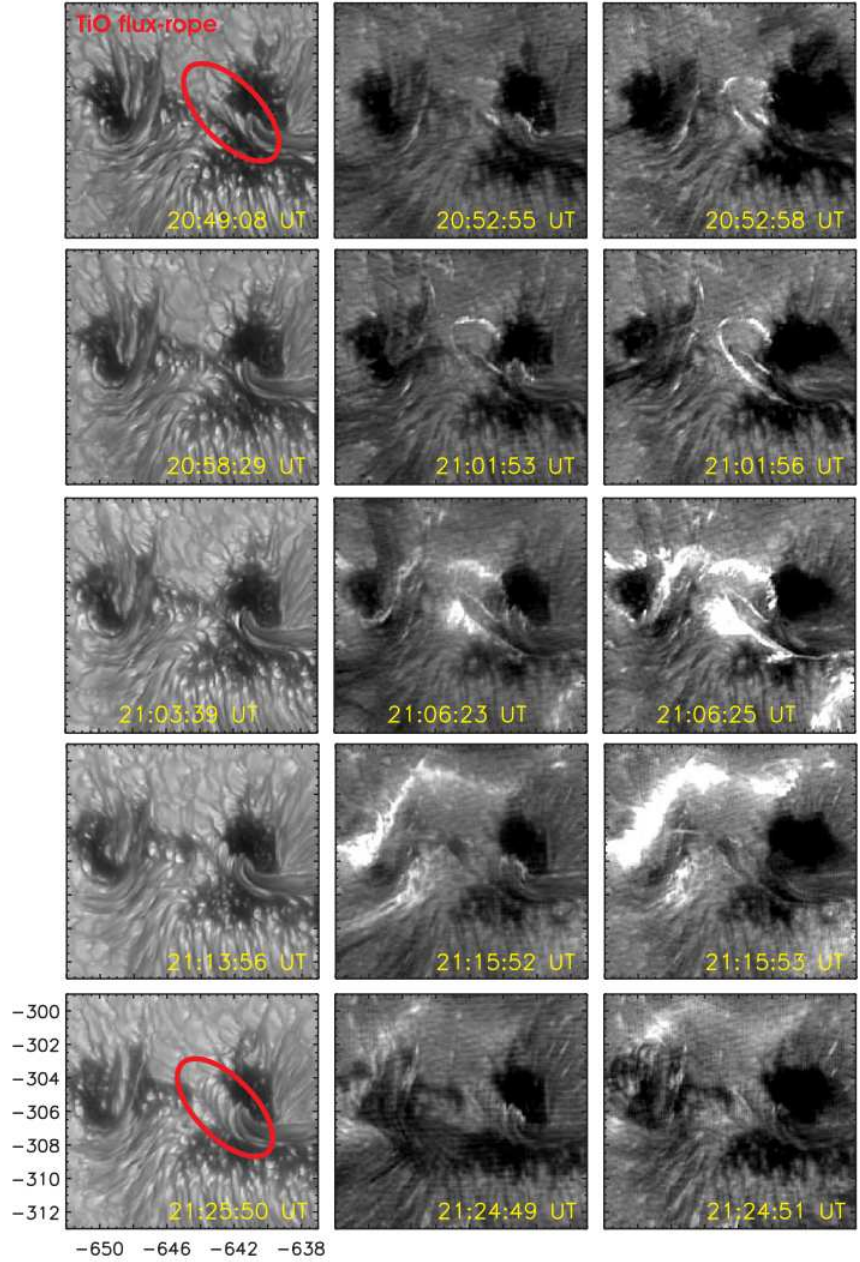}
\caption{Zoomed NST images of the region with highly twisted magnetic field. TiO images are shown in the left column. The blue wing and red wing H$\rm_{\alpha}$ filtergrams are presented correspondingly in the central and right columns. The twisted TiO magnetic structure called TiO flux-rope is marked by red ellipse in the top-left and bottom-left panels.}
\label{VIS_TIO}
\end{figure}

\begin{figure}[h!]
\centering
\includegraphics[width=0.8\linewidth]{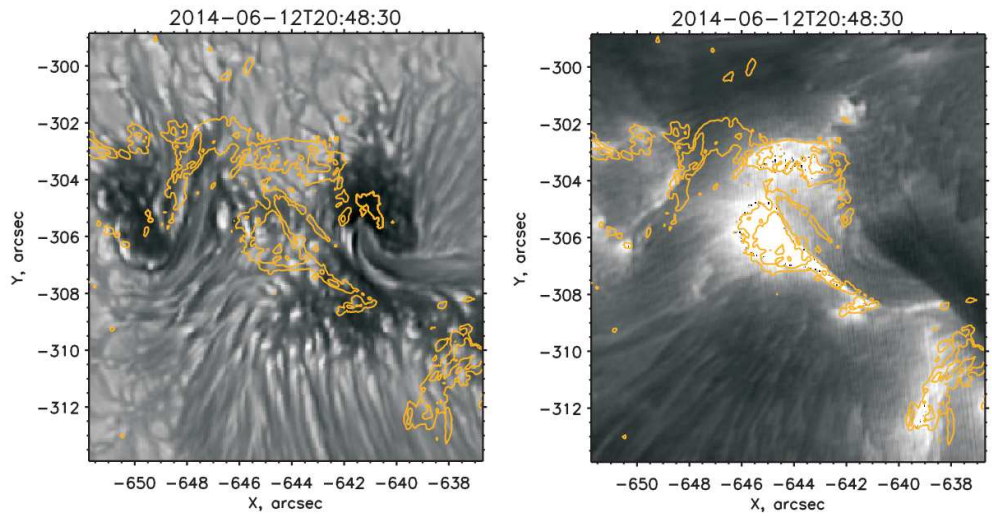}
\caption{The correspondence between the H$\rm_{\alpha}$ emission sources and the photospheric structures illustrated by overlaying the TiO and H$\rm_{\alpha}$ images. Orange contours correspond to the emission source seen in the H$\rm_{\alpha}$-line red-wing filtergrams. Background layers are the corresponding TiO image (left panel), and H$\rm_{\alpha}$ line core filtergrams (right panel).}
\label{VIS_compare}
\end{figure}

Detailed images of the flare region in the vicinity of the PIL are displayed in Fig.\,\ref{VIS_TIO}. The TiO images in the left column show the temporal dynamics of an elongated structure (that looks like a small twisted flux rope, or an arcade, and indicated by red ellipse), separating two sunspots of opposite magnetic polarity (`$\delta$-type' configuration). We will refer to this structure as the TiO arcade hereinafter. It had been described by \cite{Sadykov2015} and \cite{Kumar2015}. The visible width of this structure has a tendency to grow with time. Two other columns in Fig.\,\ref{VIS_TIO} show images in the red and blue wings of $\rm H\alpha$ line. At beginning of the flare, before the impulsive phase, we observe tiny $\rm H\alpha$ ribbons and brightenings in the vicinity of the TiO arcade. At 21:06:25 UT the most intensive $\rm H\alpha$ ribbons surround the TiO arcade (which may correspond to a magnetic flux rope). A very intriguing peculiarity is that we observe a three ribbon structure: two relatively wide ribbons (with width $\sim 1^{\prime\prime}$) and a tiny thread (width $\sim 0.3^{\prime\prime}$) located between them. The tiny ribbon directly corresponds to the TiO flux-rope which is located inside the twisted magnetic structure seen in the $\rm H\alpha$ line core. These observations are presented in Fig.\,\ref{VIS_compare} in more details, where $\rm H\alpha$ (line wing) emission sources (shown by yellow contour lines) are compared with the TiO and $\rm H\alpha$ (line core) images. The thin ribbon was stable while the others moved away from each other with the speed of $\sim 5-10$ km/s.

\subsection{DYNAMICS OF PHOTOSPHERIC PLASMA FLOWS}

In this section we present analysis of the Doppler line-of-sight (LOS) velocity in the flare region, measured by HMI. It is worth noting that the detected flows do not correspond to the upflows and downflows directly as the solar flare is located $\sim 640^{\prime\prime}$ from the disk center. Thus, the LOS velocities are composed of the horizontal (along the solar surface) and vertical components. However, we will define the positive and negative LOS velocities as `downflows' and `upflows'.

\begin{figure}[h!]
\centering
\includegraphics[width=0.7\linewidth]{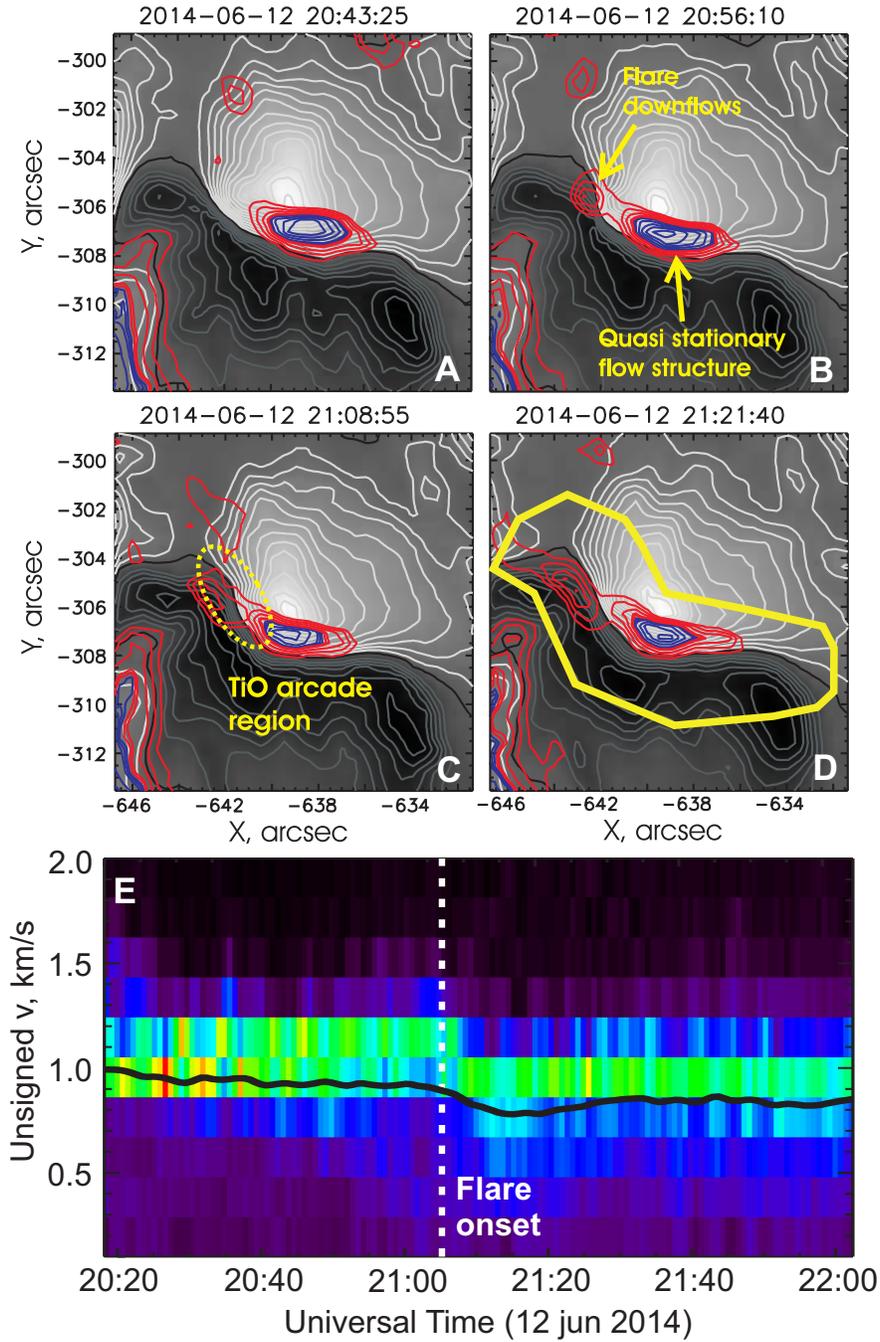}
\caption{Evolution of the line-of-sight velocity in the PIL (blue contours show upflows, red -- downflows): A-B) before the flare; C) during the flare; D) after the flare. Panel~E shows the temporal dynamics in the form of a Doppler speed histogram calculated inside the region marked by yellow contour in panel~D.}
\label{HMI_v}
\end{figure}

The Dopplergrams are compared with the LOS magnetograms in Fig.\,\ref{HMI_v}A-D. Before the flare onset (Fig.\,\ref{HMI_v}A) we observe a system of upflows surrounded by downflows near the PIL. This flow system was observed during the whole flare process without significant changes. The flare initiation is associated with appearance of new downflows (Fig.\,\ref{HMI_v}B-D) near another region of the PIL (marked by an arrow) with \textbf{a speed value up to 1~km/s}, which corresponds to the TiO arcade.

In Fig.\,\ref{HMI_v}E, we demonstrate the dynamics of the flow speed distribution. Colors from black to red indicate the number of the HMI pixels in particular speed ranges indicated along y-axis. By yellow contour in Fig.\,\ref{HMI_v}D we show an area where the distribution of Doppler speeds was calculated. This place includes the TiO arcade and a surrounding region with enhanced Doppler flows. One can notice that after the flare onset (marked by vertical line in panel E) the distribution of the Doppler speeds experiences changes. Thick black curve is the Doppler speed ($\sim 1$ km/s) averaged over all pixels inside the yellow contour in Fig.\,\ref{HMI_v}D. This line also indicates changing averaged flow speed \textbf{(from 1 to 0.75~m/s)} in the region of our interest, associated with the flare onset and marked by dashed white line in Fig.\,\ref{HMI_v}E.

\subsection{ELECTRIC CURRENTS IN THE FLARE REGION}

To calculate vertical currents, $j_z$ (component perpendicular to the solar surface), we use the HMI vector magnetograms, and the Ampere's law \citep[e.g.][]{Guo2013,Musset2015}:
\vskip-.6cm
\begin{eqnarray}
j_z=\frac{c}{4\pi}(\nabla\times\vec{B})_z = \frac{c}{4\pi}\left(\frac{\partial B_x}{\partial y} - \frac{\partial B_y}{\partial x}\right)
\label{eq_jz}
\end{eqnarray}
where $B_x$ and $B_y$ correspond to the East-West and South-North components of magnetic field.

\begin{figure}[h!]
\centering
\includegraphics[width=1.00\linewidth]{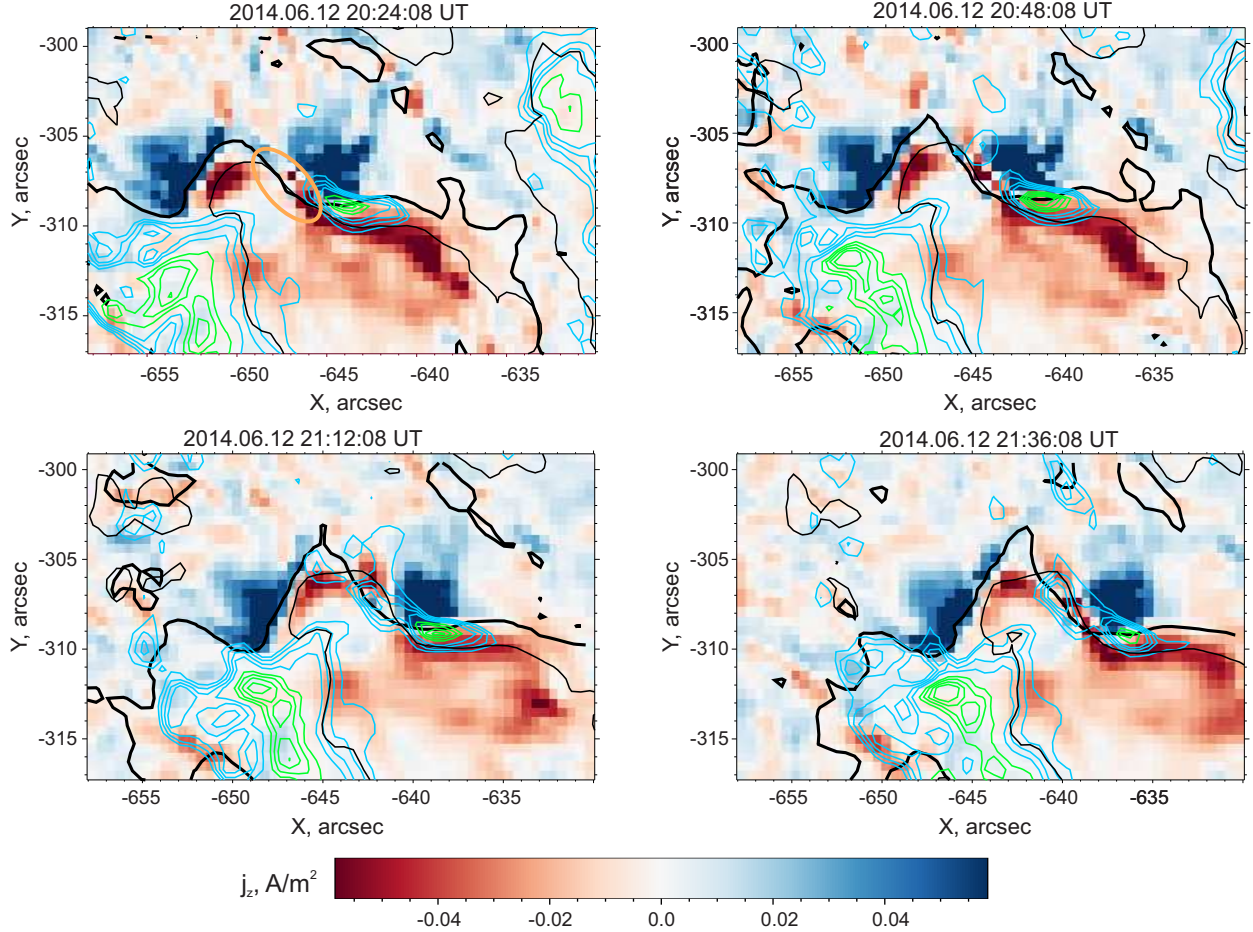}
\caption{Sequence of 4 red-blue images showing the evolution of the vertical component of electric current density. Green and cyan contours show upflows (-0.8, -0.6, -0.4 and -0.2 m/s) and downflows (0.2, 0.4, 0.6, 0.8 km/s) according to the HMI Doppler measurements. Thin black contours show the LOS PIL, while the thick black lines mark the reconstructed PIL reconstructed from the HMI vector magnetograms. Orange ellipse marks the region of TiO arcade.}
\label{jz}
\end{figure}

Since the flare was located far from the disk center the observed LOS PIL may deviate from the exact position of the PIL of the $B_z$-component (vertical to the solar surface) due to the projection effect. Using the HMI vector magnetograms we recalculated all $\vec{B}$ components from the local helioprojective Cartesian system to the Heliocentric Spherical coordinate system. The recalculated values are used for calculations of $j_z$ shown in Fig.\,\ref{jz}. The figure also shows comparison of the LOS PIL (thin black line) with the exact PIL (thick black line) in the helioprojective Cartesian coordinate system.

\begin{figure}[h!]
\centering
\includegraphics[width=1.0\linewidth]{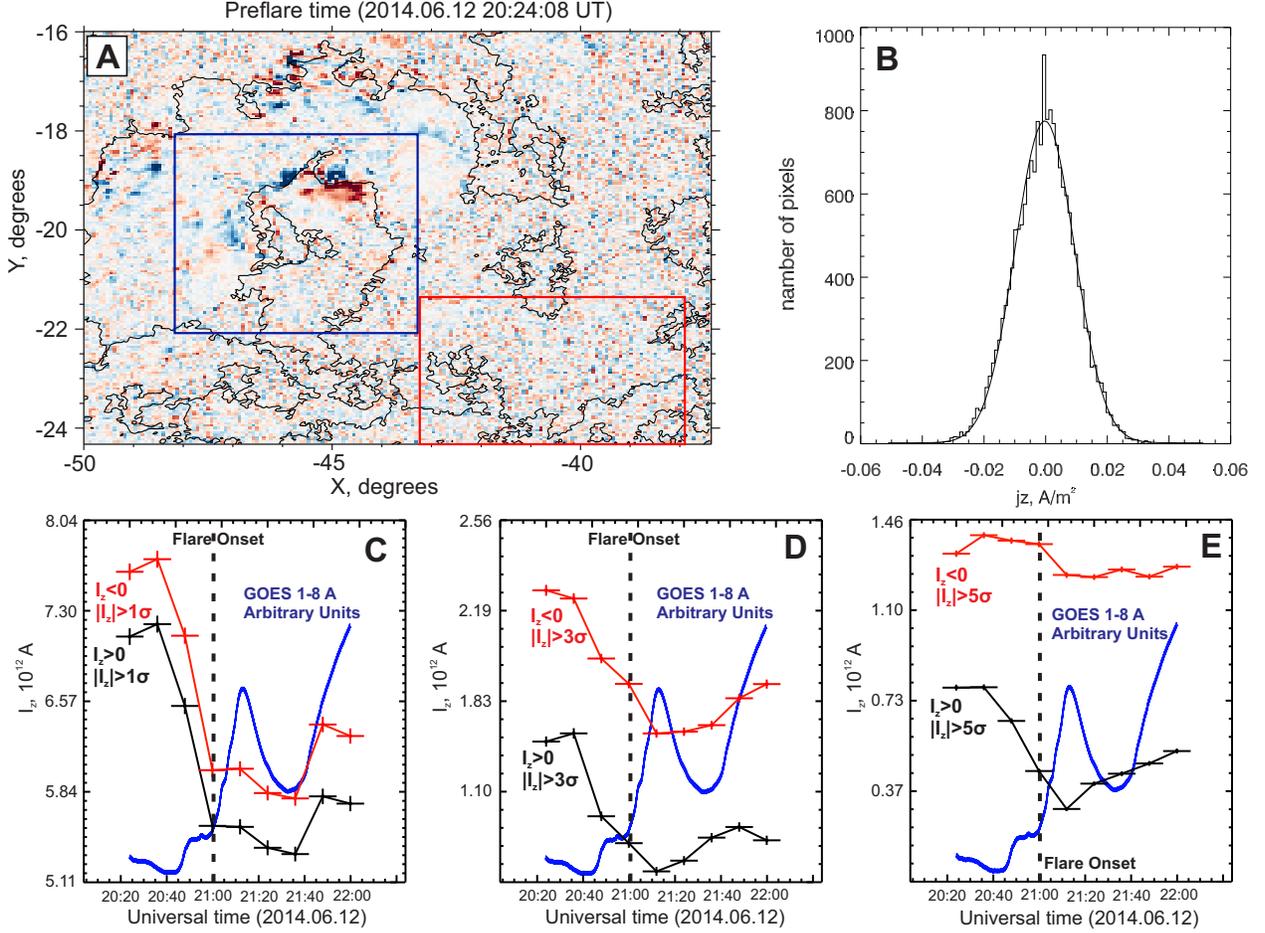}
\caption{Temporal dynamics of total vertical electric currents $I_z$ in the flare region (marked by blue rectangle in panel A) is shown in panels C, D and E. These three panels show the total currents above thresholds with values $1\sigma$, $3\sigma$ and $5\sigma$. Red rectangle in panel A corresponds to the non-flaring region where we calculate the distribution of electric currents to determine the noise level $\sigma$. This distribution is shown in panel B by histogram, where solid line is a Gaussian fit. Blue curve in panels C and E is the GOES X-ray lightcurve (1-8~\AA) in arbitrary units.}
\label{jz_evo}
\end{figure}

There are several places along the PIL where electric currents are intensified. In the region of the TiO flux-rope electric currents are intensive and experience changes that can be visually detected in Fig.\,\ref{jz}. To quantify the changes we calculate the total electric current through the region shown as blue box  in Fig.\,\ref{jz_evo}A. The temporal dynamics of the total $I_z$ is shown in Fig.\,\ref{jz_evo} in panels C, D and E. We present the total positive (black) and negative (red) currents, calculated above three threshold values: $1\sigma$, $3\sigma$ and $5\sigma$, where $\sigma$ is the standard deviation of the $j_z$ distribution fitted by a Gaussian (see Fig.\,\ref{jz_evo}B) outside the flare region (red rectangle area in Fig.\,\ref{jz_evo}A) chosen to characterize noise in the $j_z$ measurements. One can note that the total currents changes during the flare. The absolute negative and positive $I_z$ values for all three cases decrease just before the X-ray impulse. This can be interpreted as dissipation of electric currents resulting in the flare energy release. The decrease of $I_z$ was the largest for the $1\sigma$ threshold value, and was about $1.6\times10^{12}$ A for both, the negative and positive components. The smallest value of the $I_z$ reduction was about $0.5\times10^{12}$~A for the $5\sigma$ threshold. Thus, the change of the total electric current is mostly due to changes of relatively weak electric currents in the flare region. However the largest relative change of total electric currents was for the $5\sigma$ threshold $\Delta I/I_{max}\approx 0.63$ while this value for the $1\sigma$ threshold was 0.21. One can also note that $I_z$ begins to increase during the growth of GOES X-ray lightcurve, caused by a subsequent flare occurred in the same region.


\subsection{TOPOLOGY OF THE MAGNETIC FIELD}

\begin{figure}[h!]
\centering
\includegraphics[width=0.8\linewidth]{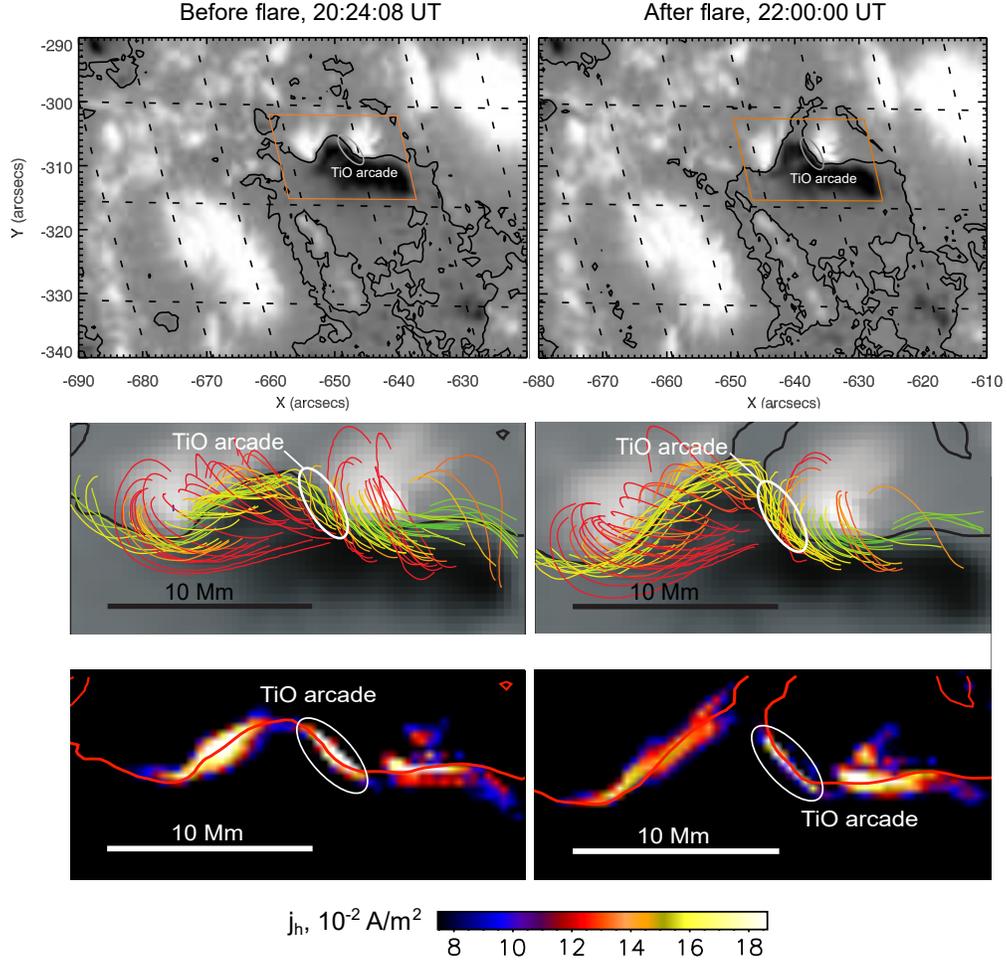}
\caption{Upper panels show the original HMI magnetograms (before and after flare) in the solar disk coordinates. Orange box indicates the region of interest selected for further analysis on the heliographic grid, and shown in middle and bottom panels. The twisted magnetic structure (result of the NLFFF extrapolation) elongated along the PIL is shown in middle panels. Different colors mark different heights of the field line (the height increases from green to red color).The left and right panels correspond to the preflare and postflare times. Bottom panels present maps of horizontal component $j_h$ of the electric current density calculated from the NLFFF extrapolations. Black line in top panels and red lines in bottom panels mark the PIL.}
\label{topology}
\end{figure}

In this section we presents results of analysis of the magnetic field structure using the NLFFF magnetic field extrapolations. Our interest is concentrated in the region surrounding the photospheric PIL. We consider the HMI magnetograms that are projected in the Heliographic spherical coordinate system. These magnetograms (after some spatial smoothing) are used to reconstruct magnetic field before and after the flare using the NLFFF method. The selected regions are marked by boxes in top panels of Fig.\,\ref{topology}, which show the original HMI magnetograms in the solar disk coordinates. The reconstructed magnetic field topology of the flare region (marked by red box in top panels) is shown in Fig.\,\ref{topology} (mid panels). The twisted magnetic field structure follows the PIL. The magnetic field topology does not change significantly during the flare energy release, as follows from comparison of the preflare and postflare NLFFF extrapolations.

Bottom panels of Fig.\,\ref{topology} show the absolute value of the horizontal component of electric current density calculated from the extrapolated magnetic field. We see that the strongest horizontal currents with values up to 0.2 A/m$^2$ are concentrated in the vicinity of the PIL. One can notice that the TiO arcade corresponds to the region of strong $j_h$. The preflare value of $j_h$ is weaker than the postflare value.

\begin{figure}[h!]
\centering
\includegraphics[width=1.0\linewidth]{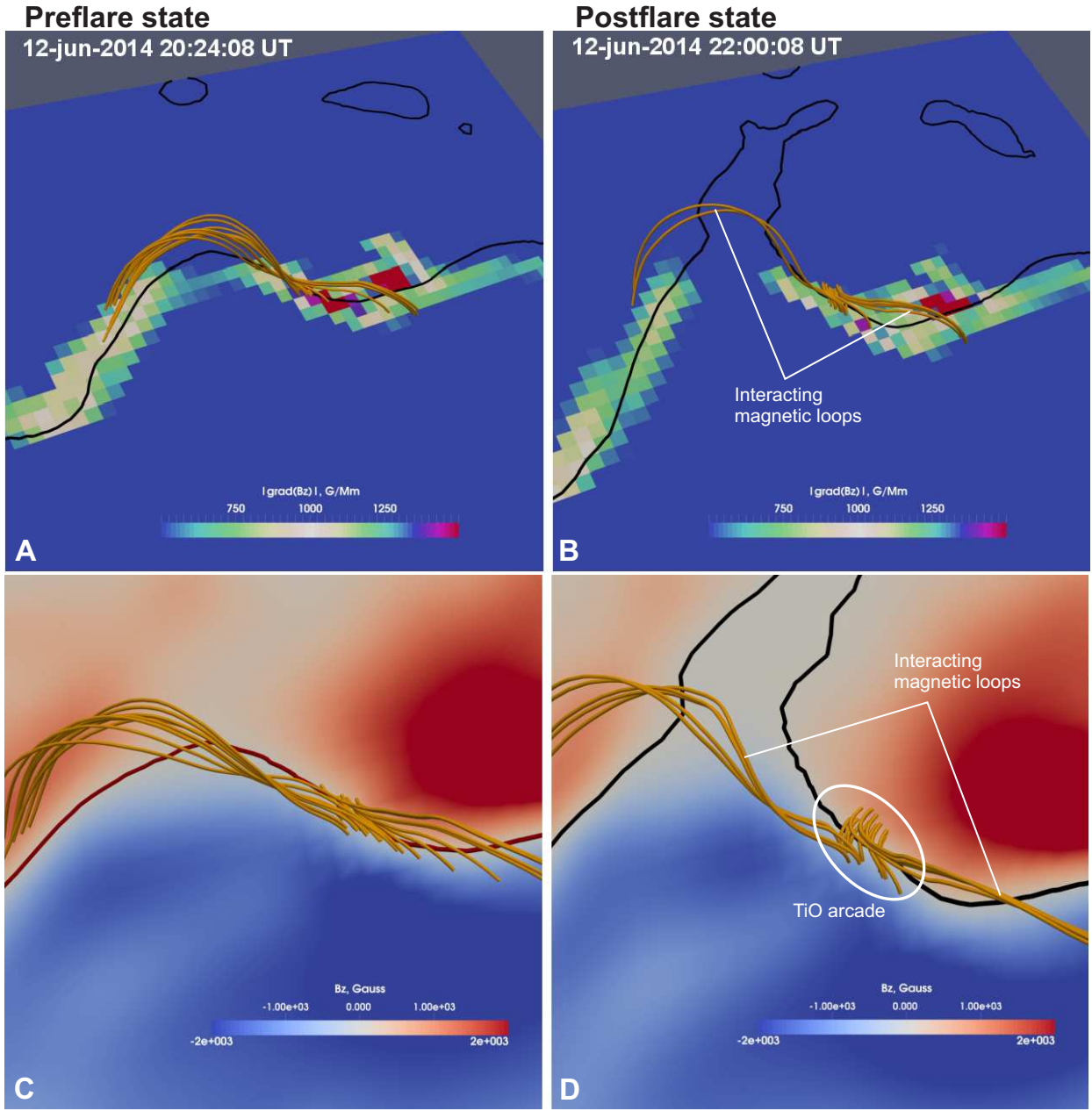}
\caption{Results of the magnetic field extrapolation in the region of the TiO flux-rope. The upper panel is a view of magnetic field lines with the background image of $\nabla_hB_z$. The bottom panels show zoomed images of the same region with view from the top, where the background layer is the photospheric magnetogram.}
\label{nlfff2}
\end{figure}

The place of our particular interest is the region where the TiO arcade and preflare activity were observed (this place is marked by ellipse in Fig.\,\ref{topology}). In Fig.\,\ref{nlfff2} we present the magnetic field extrapolation results for this region. The starting points of the magnetic field lines are selected in the region of the TiO arcade. Before the flare, we see two interacting twisted flux-tubes near the PIL. Such magnetic configuration is favourable for magnetic reconnection. The approximate place of the interaction of the magnetic flux tubes is localized at heights up to $\sim$2~Mm above the photosphere, that is in the upper chromosphere. After the flare, a small arcade is formed between these flux-tubes, probably due to magnetic reconnection. The next section describes this scenario of magnetic reconnection in more details. The results of the NLFFF modelling are consistent with the TiO and $\rm H\alpha$ images (Fig.\,\ref{VIS_TIO}). In Fig.\,\ref{nlfff2}A,B we show the magnetic field gradient $\nabla_hB_z=\sqrt{(\partial B_z/\partial x)^2+(\partial B_z/\partial y)^2}$. The strongest value of $\nabla_hB_z$ ($\sim 1-2$~kG/Mm) is found along the PIL, and concentrated near the region where the TiO flux-rope with downflows was observed. One can notice that the area of enhanced $\nabla_hB_z$ was reduced after the flare (from seven HMI pixels to one pixel where $\nabla_hB_z\gtrsim 1.3$~kG/Mm).

\begin{figure}[h!]
\centering
\includegraphics[width=0.7\linewidth]{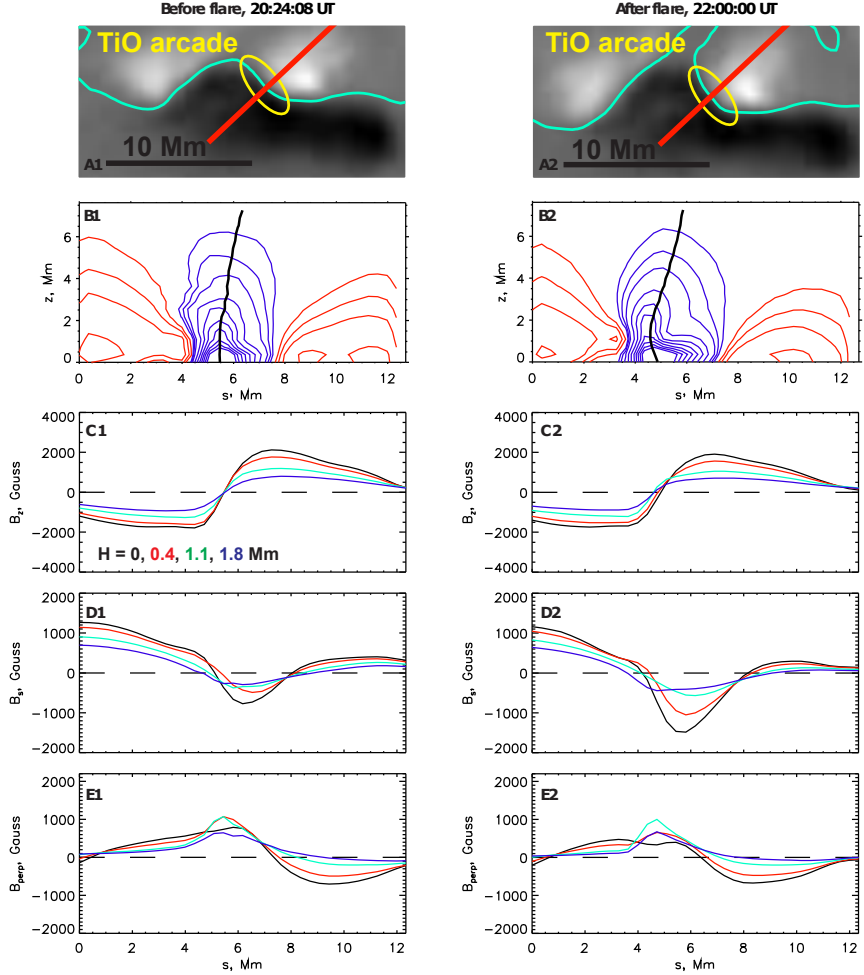}
\caption{Distribution of electric current density and magnetic field components derived from the NLFFF extrapolations along the slice marked by red line presented in panels A. The background is $B_z$ map and blue line is the PIL. Left and right columns correspond to the preflare (20:24:08~UT) and postflare (22:00:0~UT) times. Panels B show $\vec{j}$ component perpendicular to the slice (black line is the PIL in the plane of slice). Blue contours correspond to positive (directed to the reader) $j_{perp}$ with levels 6, 11, 19, 37, 56, 75, 93, 112, 130 and 150 mA/m$^2$. Red contours correspond to negative $j_{perp}$ with levels 6, 11, 19, 37, 56 and 75 mA/m$^2$. Panels C, D, and E show the following components of magnetic field: $B_z$ (vertical), $B_s$ (along the slice) and $B_{perp}$ (perpendicular to slice) along the slice at different heights: 0 (black), 0.4 (red), 1.1 (light blue), and 1.8 Mm (dark blue).}
\label{slice}
\end{figure}

It is important to trace changes in the magnetic field strength and the electric current density in the vicinity of the PIL. In Figure~\ref{slice} we demonstrate the distribution of different physical quantities derived from the NLFFF magnetic extrapolations along the slice intersecting the PIL in the vicinity of the TiO arcade. We show these distributions for preflare (left panels) and postflare (right panels) times.

The magnetic field strength at the chromospheric heights reaches $\sim 2$ kG. The strength of magnetic field components along the slice is shown by different colors. We found that the $B_s$ component increase from 800 to 1500 G at the photospheric level, and from 400 G to 600 G at the height of 1.1~Mm above the photosphere. We also show the sign inversion line of $B_z$ component in the plane of the slice by black line (panels B and C). This line experiences some changes in the region of the TiO arcade.

\begin{figure}[h!]
\centering
\includegraphics[width=0.7\linewidth]{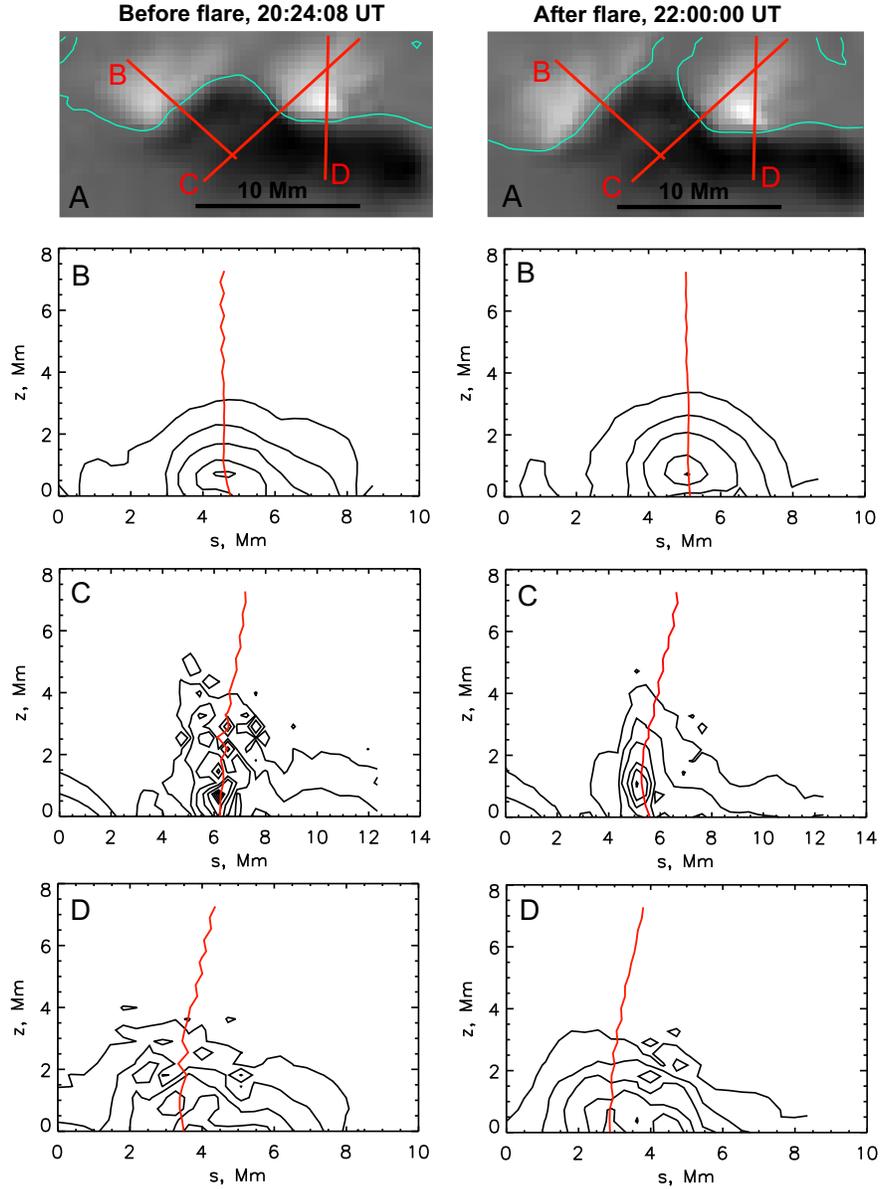}
\caption{The distribution of the absolute value of $\vec{j}$ along slices B, C, D shown in two top panels with $B_z$ map at before the flare (left panels) and after the flare (right panels) times. Panels B, C and D correspond to slices B, C and D. Contours mark electric current density levels with values: 15, 37, 75, 150, 224, 298, 336 and 373 mA/m$^2$. Red line is the PIL in the plane of the slices.}
\label{jslices}
\end{figure}

There are no significant changes in the distribution of the electric current component $j_{perp}$ perpendicular to the slice shown in panels B1 and B2 of Figure~\ref{slice}. To trace changes of the electric current density we present Figure~\ref{jslices}. Here we show three slices intersecting the PIL in different places. The strongest electric currents (with current density greater than 0.3~A/m$^2$) were concentrated in the area less than 1 Mm$^2$. One can notice that in the region of the TiO arcade (Slice C) the magnitude of the electric current density is decreased from 0.37 to 0.3~A/m$^2$, and that the height of the maximal current density is also decreased, while the other slices do not show significant changes. Thus, this analysis confirms that the changes of the electric current density are associated with a small region near the PIL, where we observe the TiO arcade. The decreasing electric currents can be explained by their dissipation during the magnetic reconnection. It is worth noting that we calculated the electric current density on the scale of the HMI resolution ($\sim 1$~arcsec/pixel), and, thus, the obtained $j_z$ value represents a low limit. In reality, the $j_z$ magnitude can be greater due to unresolved fine structures of the electric current.

\section{FLARE ENERGETICS}

We found that the flare energy release was developed near the PIL where magnetic field is highly twisted and strong electric currents are concentrated. To show that the flare energy was stored in the PIL region one need to estimate the flare energetics and compare it with the released magnetic energy in the PIL.

The studied solar flare is not eruptive and, thus, there is no CME energy in the total flare energetics. To calculate the total flare energetics one can integrate the flare emissions. The total integrated soft X-ray radiation losses are estimated as $L_{rad}^{GOES}\sim 2\times10^{29}$ ergs according to the GOES data. We use the data from https://solarflare.njit.edu \citep{Sadykov2017} where the background subtraction method of \cite{Ryan2012} was applied. This energy was estimated by time integration of $L_{rad}=EM\cdot f(T)$, where $f(T)$ is the radiative loss function \citep{Rosner1978}. Emission measure $EM$ and plasma temperature $T$ are calculated according to the method described in the work of \cite{Thomas1985}. A significant part of the flare radiation is also in the UV and EUV parts of the electromagnetic spectrum \citep{Emslie2012,Milligan2014}. We applied standard DEM-analysis \citep{Aschwanden2015} to estimate the radiation losses and the maximum thermal energy using the AIA observations. This analysis covers plasma temperatures from $10^5$ to $10^8$~K. The estimated radiation losses are $L_{rad}^{DEM}=7\times 10^{27}$~ergs which is smaller than $L_{rad}^{GOES}$.

The total internal plasma energy is calculated as $U_{th}=3k_{B}T\sqrt{EM\cdot V}$ where $k_B$ is Boltzman constant and $V$ is the plasma emitting volume which is estimated from the X-ray or EUV images depending on the T and EM estimation. For the GOES data and the flare volume ($\sim 10^{27}$~cm$^{3}$) estimated in the work of \cite{Sadykov2015} the maximum thermal energy is $U_{th}^{GOES}\approx 3\times 10^{29}$~ergs. To estimate the volume of EUV emitting plasma $V\approx 4\times 10^{27}$~cm$^{3}$ we also used the AIA images. The DEM analysis results in maximum $U_{th}^{DEM}\approx 2.4\times 10^{30}$~ergs which is much smaller than $U_{th}^{GOES}$, as in the work of \cite{Aschwanden2015} for the flares.

To estimate the total energy of nonthermal electrons we use results from the work of \cite{Sadykov2015}. Taking the spectral index of nonthermal electrons as $\delta = 7.5$ and total flux $F(E>E_{low})=3\times10^{35}$~electrons/s ($E_{low}$=13~keV) one can determine nonthermal energy $P_{nonth}\approx\Delta tFE_{low}(\delta-1)/(\delta-2)\approx 2\times10^{29}$~erg/s where $\Delta t\approx 30$~s is a characteristic duration of the HXR pulse above 12~keV.

To calculate dissipation of the energy stored in the magnetic fields generated by local electric currents one can estimate free energy which is $E_{free}=E_{nlfff}-E_{pot}$, where $E_{nlfff}$ is the total energy of magnetic field extrapolated from the HMI vector magnetogram data using the NLFFF technique, and $E_pot$ is the corresponding energy of potential (no electric currents) magnetic field with the minimal possible energy. The total magnetic energy is calculated as integration of $B^2/8\pi$ over the volume shown in the top two panels in Fig.\,\ref{topology} with height of 10~Mm. For the preflare state $E_{nlfff}=4.87\times10^{31}$~ergs, $E_{pot}=6.27\times 10^{30}$~ergs and $E_{free}=4.24\times 10^{31}$~ergs. After the flare the magnetic energies were the following: $E_{nlfff}=4.66\times10^{31}$~ergs, $E_{pot}=6.69\times 10^{30}$~ergs and $E_{free}=3.99\times 10^{31}$~ergs. Thus, the change of the free magnetic energy in the PIL region was $\Delta E_{free}=2.48\times 10^{30}$~ergs, enough to explain the total flare energetics.

\section{DISCUSSION AND CONCLUSIONS}

In this section we will try to draw a picture of the physical processes in the flare region. First of all, we summarize the main observational results. All preflare activity in the low solar atmosphere was located near the PIL in a compact region, according to the NST data. The flare observed in the NST images was accompanied by changing magnetic fields near the PIL which is clearly seen in the HMI vector magnetograms. The TiO and H$\rm_{\alpha}$~images reveal formation of a small ($\sim 3$~Mm) arcade-like magnetic structure in the photospheric and chromospheric layers of the solar atmosphere. Magnetic field of this structure is approximately $1000$ Gauss. The H$\rm_{\alpha}$ emission sources near this place were observed as a three ribbon structure with the thinnest ribbon (seen only in the wings of H$\rm_{\alpha}$ line) located between the thicker ones. This thin ribbon was stationary while the others moved away from each other during the flare with the speed $\sim 5-10$ km/s.

In the same region of the small magnetic arcade we found intensification of Doppler downflows across the PIL, which was associated with the flare onset. The strongest intensification of electric currents estimated from the HMI vector magnetograms is also located near the PIL. The observed flows near the PIL are connected with the flare energy release as their distribution and average speed experiences changes during the flare time. The flows could be triggered by magnetic reconnection as the plasma attached to magnetic field lines in the arcade moves downward.

\begin{figure}[h!]
\centering
\includegraphics[width=1.0\linewidth]{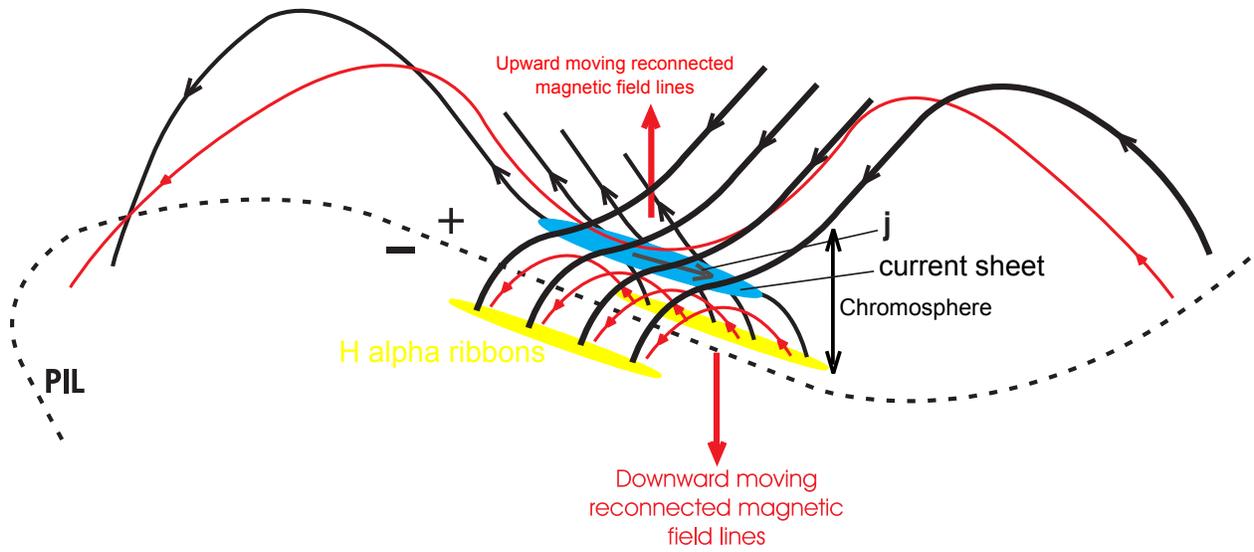}
\caption{Schematic illustration of the magnetic reconnection process in the chromosphere, which probably takes place in the studied flare. The chromospheric layer is indicated by black arrows. Red curves correspond to the reconnected magnetic field lines.}
\label{scheme}
\end{figure}

The NLFFF modelling reveals interaction of oppositely directed magnetic flux-tubes in the PIL. The strong electric current was concentrated near the PIL, and its strongest value is achieved in the region of the TiO arcade. Before the flare onset the interaction of the magnetic fluxes extended from the photospheric up to the chromospheric layers. These two interacting magnetic flux tubes in the low layers of the solar atmosphere are observed as the TiO arcade, that is in accordance with the NST observations. To explain results of the NLFFF modelling and the NST observation we present a possible scheme of the magnetic reconnection and magnetic field topology in the PIL, which is illustrated in Fig.\,\ref{scheme}. The reconnection is developed in the chromospheric region with the strong magnetic field codirectional to the electric current. Footpoints of the formed arcade (which is seen as the arcade in the NST/TiO images) may correspond to the observed H$\rm_{\alpha}$ ribbons. How can we explain the appearance of the third thin ribbon located between the thick ones in the low solar atmosphere, elongated along the PIL, and cospatial with the TiO arcade? This H$\rm_{\alpha}$ emission could originate from a thin channel (possibly a part of the current sheet) where electric current dissipates and heats the plasma causing its emission.

In Introduction we mentioned an EUV jet in the region of the TiO arcade, which was discussed in the papers of \cite{Sadykov2015} and \cite{Kumar2015}. In the framework of the discussed model this jet can be associated with the reconnected magnetic field lines and ejected plasma moving in the upward direction.

How can we explain formation of the large scale emission ribbons discussed by \cite{Kumar2015} and \cite{Sadykov2015}? Can the energy be transferred from the site in the PIL in the lower atmosphere to the flare ribbons? Perhaps, it is possible to organise the energy transfer by accelerated particles spreading along the magnetic field lines, and filling the large-scale magnetic field structure. Heat conduction fluxes can be also responsible for the energy transfer. The observations and NLFFF modelling presented in the paper of \cite{Sadykov2016} show that the flare ribbon emission source are connected with the PIL region by low-lying ($\lesssim 4.5$~Mm) magnetic field lines which could serve as channels for energy transfer from the primary energy release site.

Similar magnetic field topologies of the magnetic reconnection as in Fig.\,\ref{scheme} have been discussed previously \citep[e.g.][]{Priest2002}. The main feature of our scheme is that the magnetic reconnection in this flare develops in low layers of the solar atmosphere with a strong magnetic field component along the PIL in the current sheet between two interacting flux ropes.



Presented observational results evidence in favour of location of the primary energy release site in the chromosphere where plasma is partially ionized, in the region of strong electric currents concentrated near the polarity inversion line. Magnetic reconnection may be triggered by two interacting magnetic flux tubes with forming current sheet elongated along the PIL, resembling the classical model of \cite{Gold1960}, where authors considered two interacting twisted magnetic tubes. The reconnection process develops in the presence of strong magnetic field $\sim 1000$ Gauss and, probably, large radiative losses due to the high density. The observed small magnetic arcade can be a result of the chromospheric magnetic reconnection.

The studied event is a good illustration of a chromospheric non-eruptive flare which does not fit into a standard model. It motivates us to investigate magnetic reconnection in the physical conditions far from those that are assumed in the frame of the standard model. To develop the idea of the chromospheric magnetic reconnection we need new detailed numerical simulations whose results can be compared with high-resolution multiwavelength observations of solar flares.

The main conclusion of the paper is that the flare primary energy release develops within a relatively small volume in the vicinity of the magnetic field polarity inversion line in the presence of strong electric currents, plasma flows, magnetic field values and its gradients. It is likely that all large scale processes of flare energy release observed as UV, EUV and $\rm H\alpha$ ribbons and formation of hot UV and soft X-ray loops are connected with the small scale primary energy release site located near the PIL.

\acknowledgements

This work was partially supported by the Russian Foundation for Basic Research (grants 15-32-21078 and 16-32-00462), and by NASA grant NNX14AB68G. The authors acknowledge the BBSO observing and technical team for their contribution and support. The BBSO operation is supported by NJIT, US NSF AGS-1250818, and NASA NNX13AG14G grants, and the NST operation is partly supported by the Korea Astronomy and Space Science Institute and Seoul National University and by the strategic priority research program of CAS with Grant No. XDB09000000.


\bibliographystyle{apj}

\begin{thebibliography}{48}
\expandafter\ifx\csname natexlab\endcsname\relax\def\natexlab#1{#1}\fi

\bibitem[{{Aschwanden} {et~al.}(2015){Aschwanden}, {Boerner}, {Ryan}, {Caspi},
  {McTiernan}, \& {Warren}}]{Aschwanden2015}
{Aschwanden}, M.~J., {Boerner}, P., {Ryan}, D., {Caspi}, A., {McTiernan},
  J.~M., \& {Warren}, H.~P. 2015, \apj, 802, 53

\bibitem[{{Carmichael}(1964)}]{Carmichael1964}
{Carmichael}, H. 1964, NASA Special Publication, 50, 451

\bibitem[{{Centeno} {et~al.}(2014){Centeno}, {Schou}, {Hayashi}, {Norton},
  {Hoeksema}, {Liu}, {Leka}, \& {Barnes}}]{Centeno2014}
{Centeno}, R., {Schou}, J., {Hayashi}, K., {Norton}, A., {Hoeksema}, J.~T.,
  {Liu}, Y., {Leka}, K.~D., \& {Barnes}, G. 2014, \solphys, 289, 3531

\bibitem[{{Chae} {et~al.}(2003){Chae}, {Moon}, \& {Park}}]{Chae2003}
{Chae}, J., {Moon}, Y.-J., \& {Park}, S.-Y. 2003, Journal of Korean
  Astronomical Society, 36, 13

\bibitem[{{De Pontieu} {et~al.}(2014){De Pontieu}, {Title}, {Lemen}, {Kushner},
  {Akin}, {Allard}, {Berger}, {Boerner}, {Cheung}, {Chou}, {Drake}, {Duncan},
  {Freeland}, {Heyman}, {Hoffman}, {Hurlburt}, {Lindgren}, {Mathur}, {Rehse},
  {Sabolish}, {Seguin}, {Schrijver}, {Tarbell}, {W{\"u}lser}, {Wolfson},
  {Yanari}, {Mudge}, {Nguyen-Phuc}, {Timmons}, {van Bezooijen}, {Weingrod},
  {Brookner}, {Butcher}, {Dougherty}, {Eder}, {Knagenhjelm}, {Larsen},
  {Mansir}, {Phan}, {Boyle}, {Cheimets}, {DeLuca}, {Golub}, {Gates}, {Hertz},
  {McKillop}, {Park}, {Perry}, {Podgorski}, {Reeves}, {Saar}, {Testa}, {Tian},
  {Weber}, {Dunn}, {Eccles}, {Jaeggli}, {Kankelborg}, {Mashburn}, {Pust},
  {Springer}, {Carvalho}, {Kleint}, {Marmie}, {Mazmanian}, {Pereira}, {Sawyer},
  {Strong}, {Worden}, {Carlsson}, {Hansteen}, {Leenaarts}, {Wiesmann},
  {Aloise}, {Chu}, {Bush}, {Scherrer}, {Brekke}, {Martinez-Sykora}, {Lites},
  {McIntosh}, {Uitenbroek}, {Okamoto}, {Gummin}, {Auker}, {Jerram}, {Pool}, \&
  {Waltham}}]{DePontieu2014}
{De Pontieu}, B., {et~al.} 2014, \solphys, 289, 2733

\bibitem[{{Emslie} {et~al.}(2012){Emslie}, {Dennis}, {Shih}, {Chamberlin},
  {Mewaldt}, {Moore}, {Share}, {Vourlidas}, \& {Welsch}}]{Emslie2012}
{Emslie}, A.~G., {et~al.} 2012, \apj, 759, 71

\bibitem[{{Fletcher} {et~al.}(2011){Fletcher}, {Dennis}, {Hudson}, {Krucker},
  {Phillips}, {Veronig}, {Battaglia}, {Bone}, {Caspi}, {Chen}, {Gallagher},
  {Grigis}, {Ji}, {Liu}, {Milligan}, \& {Temmer}}]{Fletcher2011}
{Fletcher}, L., {et~al.} 2011, \ssr, 159, 19

\bibitem[{{Georgoulis} {et~al.}(2002){Georgoulis}, {Rust}, {Bernasconi}, \&
  {Schmieder}}]{Georgoulis2002}
{Georgoulis}, M.~K., {Rust}, D.~M., {Bernasconi}, P.~N., \& {Schmieder}, B.
  2002, \apj, 575, 506

\bibitem[{{Gold} \& {Hoyle}(1960)}]{Gold1960}
{Gold}, T., \& {Hoyle}, F. 1960, \mnras, 120, 89

\bibitem[{{Goode} \& {Cao}(2012)}]{Goode2012}
{Goode}, P.~R., \& {Cao}, W. 2012, in Astronomical Society of the Pacific
  Conference Series, Vol. 463, Second ATST-EAST Meeting: Magnetic Fields from
  the Photosphere to the Corona., ed. T.~R. {Rimmele}, A.~{Tritschler},
  F.~{W{\"o}ger}, M.~{Collados Vera}, H.~{Socas-Navarro}, R.~{Schlichenmaier},
  M.~{Carlsson}, T.~{Berger}, A.~{Cadavid}, P.~R. {Gilbert}, P.~R. {Goode}, \&
  M.~{Kn{\"o}lker}, 357

\bibitem[{{Guo} {et~al.}(2013){Guo}, {D{\'e}moulin}, {Schmieder}, {Ding},
  {Vargas Dom{\'{\i}}nguez}, \& {Liu}}]{Guo2013}
{Guo}, Y., {D{\'e}moulin}, P., {Schmieder}, B., {Ding}, M.~D., {Vargas
  Dom{\'{\i}}nguez}, S., \& {Liu}, Y. 2013, \aap, 555, A19

\bibitem[{{Hirayama}(1974)}]{Hirayama1974}
{Hirayama}, T. 1974, \solphys, 34, 323

\bibitem[{{Inoue} {et~al.}(2016){Inoue}, {Hayashi}, \& {Kusano}}]{Inoue2016}
{Inoue}, S., {Hayashi}, K., \& {Kusano}, K. 2016, \apj, 818, 168

\bibitem[{{Kosovichev} \& {Zharkova}(2001)}]{Kosovichev2001}
{Kosovichev}, A.~G., \& {Zharkova}, V.~V. 2001, \apjl, 550, L105

\bibitem[{{Krucker} {et~al.}(2008){Krucker}, {Battaglia}, {Cargill},
  {Fletcher}, {Hudson}, {MacKinnon}, {Masuda}, {Sui}, {Tomczak}, {Veronig},
  {Vlahos}, \& {White}}]{Krucker2008}
{Krucker}, S., {et~al.} 2008, \aapr, 16, 155

\bibitem[{{Kumar} {et~al.}(2015){Kumar}, {Yurchyshyn}, {Wang}, \&
  {Cho}}]{Kumar2015}
{Kumar}, P., {Yurchyshyn}, V., {Wang}, H., \& {Cho}, K.-S. 2015, \apj, 809, 83

\bibitem[{{Leake} {et~al.}(2013){Leake}, {Lukin}, \& {Linton}}]{Leake2013}
{Leake}, J.~E., {Lukin}, V.~S., \& {Linton}, M.~G. 2013, Physics of Plasmas,
  20, 061202

\bibitem[{{Leake} {et~al.}(2012){Leake}, {Lukin}, {Linton}, \&
  {Meier}}]{Leake2012}
{Leake}, J.~E., {Lukin}, V.~S., {Linton}, M.~G., \& {Meier}, E.~T. 2012, \apj,
  760, 109

\bibitem[{{Leake} {et~al.}(2014){Leake}, {DeVore}, {Thayer}, {Burns},
  {Crowley}, {Gilbert}, {Huba}, {Krall}, {Linton}, {Lukin}, \&
  {Wang}}]{Leake2014}
{Leake}, J.~E., {et~al.} 2014, \ssr, 184, 107

\bibitem[{{Lemen} {et~al.}(2012){Lemen}, {Title}, {Akin}, {Boerner}, {Chou},
  {Drake}, {Duncan}, {Edwards}, {Friedlaender}, {Heyman}, {Hurlburt}, {Katz},
  {Kushner}, {Levay}, {Lindgren}, {Mathur}, {McFeaters}, {Mitchell}, {Rehse},
  {Schrijver}, {Springer}, {Stern}, {Tarbell}, {Wuelser}, {Wolfson}, {Yanari},
  {Bookbinder}, {Cheimets}, {Caldwell}, {Deluca}, {Gates}, {Golub}, {Park},
  {Podgorski}, {Bush}, {Scherrer}, {Gummin}, {Smith}, {Auker}, {Jerram},
  {Pool}, {Soufli}, {Windt}, {Beardsley}, {Clapp}, {Lang}, \&
  {Waltham}}]{Lemen2012}
{Lemen}, J.~R., {et~al.} 2012, \solphys, 275, 17

\bibitem[{{Lin} {et~al.}(2002){Lin}, {Dennis}, {Hurford}, {Smith}, {Zehnder},
  {Harvey}, {Curtis}, {Pankow}, {Turin}, {Bester}, {Csillaghy}, {Lewis},
  {Madden}, {van Beek}, {Appleby}, {Raudorf}, {McTiernan}, {Ramaty}, {Schmahl},
  {Schwartz}, {Krucker}, {Abiad}, {Quinn}, {Berg}, {Hashii}, {Sterling},
  {Jackson}, {Pratt}, {Campbell}, {Malone}, {Landis}, {Barrington-Leigh},
  {Slassi-Sennou}, {Cork}, {Clark}, {Amato}, {Orwig}, {Boyle}, {Banks},
  {Shirey}, {Tolbert}, {Zarro}, {Snow}, {Thomsen}, {Henneck}, {McHedlishvili},
  {Ming}, {Fivian}, {Jordan}, {Wanner}, {Crubb}, {Preble}, {Matranga}, {Benz},
  {Hudson}, {Canfield}, {Holman}, {Crannell}, {Kosugi}, {Emslie}, {Vilmer},
  {Brown}, {Johns-Krull}, {Aschwanden}, {Metcalf}, \& {Conway}}]{Lin2002}
{Lin}, R.~P., {et~al.} 2002, \solphys, 210, 3

\bibitem[{{Liu} {et~al.}(2008){Liu}, {Petrosian}, {Dennis}, \&
  {Jiang}}]{Liu2008}
{Liu}, W., {Petrosian}, V., {Dennis}, B.~R., \& {Jiang}, Y.~W. 2008, \apj, 676,
  704

\bibitem[{{Magara} {et~al.}(1996){Magara}, {Mineshige}, {Yokoyama}, \&
  {Shibata}}]{Magara1996}
{Magara}, T., {Mineshige}, S., {Yokoyama}, T., \& {Shibata}, K. 1996, \apj,
  466, 1054

\bibitem[{{Milligan} {et~al.}(2014){Milligan}, {Kerr}, {Dennis}, {Hudson},
  {Fletcher}, {Allred}, {Chamberlin}, {Ireland}, {Mathioudakis}, \&
  {Keenan}}]{Milligan2014}
{Milligan}, R.~O., {et~al.} 2014, \apj, 793, 70

\bibitem[{{Musset} {et~al.}(2015){Musset}, {Vilmer}, \& {Bommier}}]{Musset2015}
{Musset}, S., {Vilmer}, N., \& {Bommier}, V. 2015, \aap, 580, A106

\bibitem[{{Ni} {et~al.}(2015){Ni}, {Kliem}, {Lin}, \& {Wu}}]{Ni2015}
{Ni}, L., {Kliem}, B., {Lin}, J., \& {Wu}, N. 2015, \apj, 799, 79

\bibitem[{{Pesnell} {et~al.}(2012){Pesnell}, {Thompson}, \&
  {Chamberlin}}]{Pesnell2012}
{Pesnell}, W.~D., {Thompson}, B.~J., \& {Chamberlin}, P.~C. 2012, \solphys,
  275, 3

\bibitem[{{Priest} \& {Forbes}(2000)}]{Priest2000}
{Priest}, E., \& {Forbes}, T. 2000, {Magnetic Reconnection} (Cambridge
  University Press), 612

\bibitem[{{Priest} \& {Forbes}(2002)}]{Priest2002}
{Priest}, E.~R., \& {Forbes}, T.~G. 2002, \aapr, 10, 313

\bibitem[{{Rosner} {et~al.}(1978){Rosner}, {Tucker}, \& {Vaiana}}]{Rosner1978}
{Rosner}, R., {Tucker}, W.~H., \& {Vaiana}, G.~S. 1978, \apj, 220, 643

\bibitem[{{Ryan} {et~al.}(2012){Ryan}, {Milligan}, {Gallagher}, {Dennis},
  {Tolbert}, {Schwartz}, \& {Young}}]{Ryan2012}
{Ryan}, D.~F., {Milligan}, R.~O., {Gallagher}, P.~T., {Dennis}, B.~R.,
  {Tolbert}, A.~K., {Schwartz}, R.~A., \& {Young}, C.~A. 2012, \apjs, 202, 11

\bibitem[{{Sadykov} {et~al.}(2017){Sadykov}, {Gupta}, {Kosovichev}, {Oria}, \&
  {Nita}}]{Sadykov2017}
{Sadykov}, V.~M., {Gupta}, R., {Kosovichev}, A.~G., {Oria}, V., \& {Nita},
  G.~M. 2017, arXiv:1702.02991

\bibitem[{{Sadykov} {et~al.}(2016){Sadykov}, {Kosovichev}, {Sharykin},
  {Zimovets}, \& {Vargas Dominguez}}]{Sadykov2016}
{Sadykov}, V.~M., {Kosovichev}, A.~G., {Sharykin}, I.~N., {Zimovets}, I.~V., \&
  {Vargas Dominguez}, S. 2016, \apj, 828, 4

\bibitem[{{Sadykov} {et~al.}(2014){Sadykov}, {Vargas Dominguez}, {Kosovichev},
  \& {Sharykin}}]{Sadykov2014}
{Sadykov}, V.~M., {Vargas Dominguez}, S., {Kosovichev}, A.~G., \& {Sharykin},
  I.~N. 2014, arXiv:1412.0172v1

\bibitem[{{Sadykov} {et~al.}(2015){Sadykov}, {Vargas Dominguez}, {Kosovichev},
  {Sharykin}, {Struminsky}, \& {Zimovets}}]{Sadykov2015}
{Sadykov}, V.~M., {Vargas Dominguez}, S., {Kosovichev}, A.~G., {Sharykin},
  I.~N., {Struminsky}, A.~B., \& {Zimovets}, I. 2015, \apj, 805, 167

\bibitem[{{Scherrer} {et~al.}(2012){Scherrer}, {Schou}, {Bush}, {Kosovichev},
  {Bogart}, {Hoeksema}, {Liu}, {Duvall}, {Zhao}, {Title}, {Schrijver},
  {Tarbell}, \& {Tomczyk}}]{Scherrer2012}
{Scherrer}, P.~H., {et~al.} 2012, \solphys, 275, 207

\bibitem[{{Schrijver}(2016)}]{Schrijver2016}
{Schrijver}, C.~J. 2016, \apj, 820, 103

\bibitem[{{Severnyi}(1958)}]{Severnyi1958}
{Severnyi}, A.~B. 1958, \sovast, 2, 310

\bibitem[{{Sturrock} \& {Coppi}(1966)}]{Sturrock1966}
{Sturrock}, P.~A., \& {Coppi}, B. 1966, \apj, 143, 3

\bibitem[{{Sudol} \& {Harvey}(2005)}]{Sudol2005}
{Sudol}, J.~J., \& {Harvey}, J.~W. 2005, \apj, 635, 647

\bibitem[{{Thomas} {et~al.}(1985){Thomas}, {Crannell}, \& {Starr}}]{Thomas1985}
{Thomas}, R.~J., {Crannell}, C.~J., \& {Starr}, R. 1985, \solphys, 95, 323

\bibitem[{{Tsuneta}(1997)}]{Tsuneta1997}
{Tsuneta}, S. 1997, \apj, 483, 507

\bibitem[{{Varsik} {et~al.}(2014){Varsik}, {Plymate}, {Goode}, {Kosovichev},
  {Cao}, {Coulter}, {Ahn}, {Gorceix}, \& {Shumko}}]{Varsik2014}
{Varsik}, J., {et~al.} 2014, in \procspie, Vol. 9147, Ground-based and Airborne
  Instrumentation for Astronomy V, 91475D

\bibitem[{{Wang} {et~al.}(2015){Wang}, {Cao}, {Liu}, {Xu}, {Liu}, {Zeng},
  {Chae}, \& {Ji}}]{Wang2015}
{Wang}, H., {Cao}, W., {Liu}, C., {Xu}, Y., {Liu}, R., {Zeng}, Z., {Chae}, J.,
  \& {Ji}, H. 2015, Nat. Commun., 6

\bibitem[{{Wang} {et~al.}(1994){Wang}, {Ewell}, {Zirin}, \& {Ai}}]{Wang1994}
{Wang}, H., {Ewell}, Jr., M.~W., {Zirin}, H., \& {Ai}, G. 1994, \apj, 424, 436

\bibitem[{{Welsch} \& {Li}(2008)}]{Welsch2008}
{Welsch}, B.~T., \& {Li}, Y. 2008, in Astronomical Society of the Pacific
  Conference Series, Vol. 383, Subsurface and Atmospheric Influences on Solar
  Activity, ed. R.~{Howe}, R.~W. {Komm}, K.~S. {Balasubramaniam}, \& G.~J.~D.
  {Petrie}, 429

\bibitem[{{Wheatland} {et~al.}(2000){Wheatland}, {Sturrock}, \&
  {Roumeliotis}}]{Wheatland2000}
{Wheatland}, M.~S., {Sturrock}, P.~A., \& {Roumeliotis}, G. 2000, \apj, 540,
  1150

\bibitem[{{W{\"o}ger} {et~al.}(2008){W{\"o}ger}, {von der L{\"u}he}, \&
  {Reardon}}]{Woeger2008}
{W{\"o}ger}, F., {von der L{\"u}he}, O., \& {Reardon}, K. 2008, \aap, 488, 375

\end{thebibliography}

\clearpage

\clearpage
\end{document}